\documentclass[twocolumn,showpacs,showkeys]{revtex4}
\usepackage{amsfonts}
\usepackage[T2A]{fontenc}
\usepackage[cp1251]{inputenc}
\usepackage{amsmath}
\usepackage{amssymb}
\usepackage[russian,english]{babel}
\usepackage{graphicx}

\begin{document}

\title{On the microscopic theory of the exciton ring fragmentation}

\author{A.V. Paraskevov}

\thanks{E-mail address: paraskev@kiae.ru}

\author{T.V. Khabarova}

\affiliation{Russian Research Center "Kurchatov Institute",
Kurchatov Sq. 1, Moscow 123182, Russia}

\begin{abstract}

The description is presented for the dependence of the indirect
exciton condensate density at the ring as a function of the polar
angle at zero temperature with the involvement of the processes of
formation and recombination of the excitons.  In particular,
starting from the quasi one-dimensional Gross-Pitaevskii equation
with a spatially uniform generating term, we derive an exact
analytical solution yielding the fragmentation of an exciton ring
which is probably observed in the experiments.

\end{abstract}



\maketitle

\textbf{1. Introduction.} In the recent experiments of the Butov's
group \cite{1, 2} the macroscopically ordered coherent state
(MOCS) of the indirect exciton system has been observed in the
coupled quantum wells. The state represents a periodic
fragmentation of the ring formed in the intersection domain of the
surplus densities of electrons in one well and photoexcited holes
in another \cite{3, 4, 5}. The fragmentation is observed below
some critical temperature ($T_{c}\sim1K$) and within a certain
range of the laser pumping power when the density $n$ of indirect
excitons (IEs) at the ring is sufficiently large and the excitons
are well defined ($n\sim10^{9}cm^{-2}$,
an average spacing between IEs at the ring is  $\bar{a}_{ex-ex}\sim n^{-1/2}%
\sim 0.1\mu m$, the IE Bohr radius is $a_{B}\sim10^{-2}\mu m$,
$\bar {a}_{ex-ex}/a_{B}\sim 10$). The typical parameters of the
fragmented ring are as follows \cite{2}: average ring radius is
$R\sim10^{2}\mu m$, the ring width is $\Delta R\sim10\mu m$, the
average radius of a fragment is $r_{fr}\approx\Delta R/2$, and a
number of fragments on the ring is $N_{fr}\approx50$. The evidence
for the MOCS coherence results from a clear interference pattern
of the IE recombination radiation from one of the ring fragments
\cite{6}.

At present, there exist two different phenomenological theories of
the MOCS formation. The first theory \cite{7} is based on the
assumption that the IE system represents a degenerate
two-dimensional quasi-ideal Bose gas and the rate of aggregating
an electron and a hole to form an indirect exciton is proportional
to the factor $\left( 1+N_{0}\right)$, where $N_{0}$ is the number
of the IEs in the zero momentum state. Under definite conditions
this leads to the instability against fluctuations of the IE ring
density. However, the theory seems to be intrinsically
inconsistent since, on the one hand, the dipole-dipole interaction
between the excitons is neglected, what is allowable only in the
low density limit, and, on the other hand, the exciton condensate
density should be sufficiently large so that the contribution from
the stimulated processes would be essential.
\par
The second theory \cite{8} is based on that the total contribution
of the repulsive dipole-dipole and attractive van der Waals
interactions becomes attractive at the distances smaller than
several IE radii. According to this theory, the ring fragmentation
is due to formation of the islands of electron-hole liquid and is
not associated with the Bose-Einstein condensation of the IEs. (We
also mention paper \cite{9} in which the IE energy distribution is
suggested to be the Boltzmann one and, along with an attractive
pair potential, the repulsive three-body interaction between IEs
is considered. At the same time the nature of the latter is not
explained.) However, the results of the experiment \cite{10}
unambiguously evidence for the repulsive character of the IE
interaction. This is consistent with the idea that just this
interaction results in a rapid IE thermalization (along with the
nonconservation of a transverse momentum) and in the screening of
the surface defects in the quantum wells plane.  Note that any
extension beyond the dipole approximation while considering the
interaction between the IEs is equivalent to the deviation from
the Bose statistics, which should self-consistently be taken into
account.

As is seen from the pictures of the spatial distribution of the IE
luminescence (see \cite{1, 2}), there exist two different spatial
periods on the fragmented ring, namely, the size of a fragment and
the size of a dark region separating adjacent fragments.
Therefore, it is clear that the density as a function of the polar
angle should be expressed via a periodic function with
nonsymmetric half-periods, i.e., an elliptic function. So, if
$\rho\left( \phi\right)$ is an elliptic function, it satisfies
equation $\left( \rho_{\phi }^{\prime}\right) ^{2}=P_{4}\left(
\rho\right)$, where $P_{4}\left( \rho\right)$ is a quartic
polynomial in $\rho$. To the same form one can reduce the
one-dimensional stationary Gross-Pitaevskii equation without an
external potential, which, as is well known, describes a
condensate of Bose particles. Hence, the idea arises that the ring
fragmentation can be explained assuming that under experimental
conditions an IE condensate exists on the ring. In addition, the
formation and recombination of excitons \textit{on the ring} can
be taken into account by their interaction with some scalar
generating, or "source-drain", field independent of coordinates.
Formally, this is similar to the contribution from the dipole
interaction of atoms with the electromagnetic field. If the number
of the IEs on the ring does not vary in time, a concept of energy
of the system is meaningful and one can find the stationary states
of the condensate as well as the corresponding density spatial
distributions.

In this Letter, we suggest the simplest microscopic theory of the
MOCS based on the thing that the presence of BEC in the IE system
(in spite of the BEC depletion due to repulsive exciton-exciton
interaction) as well as the influence of the processes of IE
formation and recombination on the ring result in the realization
of a stationary excited state of the condensate, which is not
spatially uniform. The indirect excitons are treated as genuine
Bose particles. In addition, we suppose that the radial profile of
the condensate density is governed by the processes which have no
direct relation to the ring fragmentation. This allows us to
consider a quasi one-dimensional problem.

\textbf{2. Polar angle dependence of the condensate density.} The
Hamiltonian for the exciton system with the source reads
\begin{align}
\hat{H} &  =\hat{H}_{J}+\int d\vec{r}\hat{\Psi}^{+}(\vec{r},t)\frac{\hat{p}^{2}}{2m}%
\hat{\Psi}(\vec{r},t)+\\ &  \frac{1}{2}\int
d\vec{r}d\vec{r}^{\prime}\hat{\Psi}^{+}(\vec{r},t)\hat
{\Psi}^{+}(\vec{r}^{\prime},t)U(\vec{r}-\vec{r}^{\prime})\hat{\Psi}(\vec
{r}^{\prime},t)\hat{\Psi}(\vec{r},t).\nonumber
\end{align}
Here $\hat{\Psi}(\vec{r},t)$ is a Bose field operator obeying the commutation relations $\left[
\hat{\Psi}(\vec{r}_{1},t),\hat{\Psi}^{+}(\vec {r}_{2},t)\right] =\delta\left(  \vec{r}_{1}-\vec{r}_{2}\right)
$, $\ \left[ \hat{\Psi},\hat{\Psi}\right]  =\left[  \hat{\Psi}^{+},\hat{\Psi }^{+}\right]  =0.$ For
simplicity, we suggest that the repulsive pair interaction is local, i.e., $U(\vec{r})=\lambda\delta\left(
\vec {r}\right)$ and $\lambda>0$. In fact, this means either the
Born approximation for the potential $U(r)$ with $\lambda=%
{\displaystyle\int} U(r)d\vec{r}$, or the gas approximation
$na_{2D}^{2}\ll1$, where $n$ is the two-dimensional IE density and
$a_{2D}=m\lambda /(4\pi\hbar^{2})$ is an exact $s$-scattering
length [11]. The contribution of the source is given by
\begin{equation}
\hat{H}_{J}=\int d\vec{r}\left(  J\left(  t\right)
\hat{\Psi}^{+}(\vec {r},t)+J^{\ast}\left(  t\right)
\hat{\Psi}(\vec{r},t)\right)  ,
\end{equation}
where $J\left(  t\right)  =\left\vert J\right\vert e^{i\alpha\left( t\right) }$. It is natural to assume that
the time dependence of the source is the same as for the wave function of the ground condensate state in the
lack of the source, $\alpha\left( t\right)  =-\mu t$. We suggest that the quantities $\left\vert J\right\vert
$ and $\mu$ are time-independent.

Let us consider the equation of motion for $\hat{\Psi}$-operator,
$i\hbar\,\partial\hat{\Psi}(\vec{r},t)/\partial t=\left[
\hat{\Psi}(\vec{r},t),\hat {H}\right]  $, assuming zero
temperature of the system. Here, the most fraction of excitons is
in the Bose-condensate state and thus we neglect noncondensate
particles. Then $\hat{\Psi}$-operator becomes the $c$-number and
obeys the equation
\begin{equation}
i\hbar\dot{\Psi}=-\frac{\hbar^{2}}{2m}\nabla^{2}\Psi+\lambda\left\vert
\Psi\right\vert ^{2}\Psi+J,\label{00}%
\end{equation}
which for $J=0$ turns into the Gross-Pitaevskii equation.
Supposing that the radial profile of the condensate density is
unvaried, we consider a quasi 1D problem. Then, the genuine
density "in amplitude" refers to the result obtained as
$n_{true}/n\sim\bar{a}_{ex-ex}/\Delta R\sim 10^{-2}$. For the
polar frame with the origin at the ring center $\nabla^{2}\approx
R^{-2}\,\partial^{2}/\partial\phi^{2}$, $d\vec{r}\approx2\pi
R\Delta Rd\phi=Sd\phi$, where $\phi$ is the polar angle, $R$ is
the ring radius, and $R\neq R\left( n\right)$. The aim of the work
is to find a qualitative behavior $n(\phi)$ at various magnitudes
of the parameters.

If one measures energy in units $E_{R}=\hbar^{2}/(2mR^{2})$, time in $\tau_{R}=\hbar /E_{R}$, and density in
$S^{-1}$, the constants in Eq.(\ref{00}) are dimensionless
\begin{equation}
i\dot{\Psi}=-\Psi_{\phi\phi}^{\prime\prime}+\lambda\left\vert
\Psi\right\vert ^{2}\Psi+J.\label{1}%
\end{equation}
Let us seek for the solution (\ref{1}) as $\Psi=\rho e^{if},$
where $\rho^{2}\approx n$ and $f$ are the density and phase of the
condensate, respectively. Then, for the real and imaginary parts
of $\Psi$ the following equations are valid
\begin{align}
\rho\dot{f}+\rho\left(  f_{\phi}^{\prime}\right) ^{2}-\rho_{\phi
\phi}^{\prime\prime}+\lambda\rho^{3}+\left\vert J\right\vert
\cos\left( \alpha-f\right)   &  =0,\\
\dot{\rho}+2\rho_{\phi}^{\prime}f_{\phi}^{\prime}+\rho f_{\phi
\phi}^{\prime\prime}-\left\vert J\right\vert \sin\left(
\alpha-f\right)
&  =0.\label{31}%
\end{align}
Multiplying both sides of Eq.(\ref{31}) by $2\rho$ and integrating
over $\phi$, we obtain that the rate of variation of the particle
number $N$ equals
\begin{equation}
\dot{N}=\int d\vec{r}2\rho\dot{\rho}=\int d\vec{r}2\rho\left\vert
J\right\vert \sin\left(  \alpha-f\right).
\end{equation}
In what follows, we are interested in the stationary currentless
states alone: $\dot{N}=0$, $f_{\phi}^{\prime}=0$. Then
$f=\alpha+\pi l$, $l$ is an integer. So,
\begin{equation}
\rho_{\phi\phi}^{\prime\prime}+\mu\rho-\lambda\rho^{3}-\left\vert
J\right\vert \left(  -1\right)  ^{l}=0,\label{2}%
\end{equation}
in which $\alpha\left(  t\right)  =-\mu t$ is taken into account.
In the case of the homogeneous density
($\rho_{\phi}^{\prime}\equiv0$, $\rho=\rho_{0}$) and $\left\vert
J\right\vert =0$, $\mu$ has a meaning of the chemical potential of
the system:
\begin{equation}
\mu=\lambda\rho_{0}^{2}.\label{3}%
\end{equation}
Note that for $\left\vert J\right\vert =0$ even in the
unhomogeneous case the quantity $\mu$ can literally be treated as
a chemical potential for the ground state of the system. In this
case it can be introduced via $\Psi(t)\sim\exp\left( -i\mu
t\right)$.

For $J\neq0$ the uniform density is determined from the equation
\begin{equation}
\mu\rho_{0}-\lambda\rho_{0}^{3}-\left\vert J\right\vert \left(
-1\right)
^{l}=0. \label{10}%
\end{equation}
Employing the Cardano formulae, one can find from the analysis
that there exist three regions in which: (i) homogeneous solution
is absent (even $l$, $\mu<\mu_{\ast}\equiv 3\lambda\left(
\left\vert J\right\vert /(2\lambda ) \right) ^{2/3}$), (ii)
homegeneous solution is unique (even $l$, $\mu=\mu_{\ast}$, and
odd $l$ at $\mu\leqslant\mu_{\ast}$), and (iii) there exist 1 or 2
various homogeneous solutions (arbitrary $l$, $\mu>\mu_{\ast}$).

For the unhomogeneous case $\rho_{\phi}^{\prime}\neq0$, the first
integral of Eq.(\ref{2}) is given by
\begin{equation}
\left(  \rho_{\phi}^{\prime}\right)  ^{2}=\frac{1}{2}\lambda\rho^{4}%
-\mu\rho^{2}+2\left\vert J\right\vert \left(  -1\right)  ^{l}\rho
+C.\label{747}%
\end{equation}
Here $C$ is a constant independent of $\phi$. Substituting
$\rho=\rho _{0}\cdot x$ and integrating, we have
\begin{equation}
\int\frac{dx}{\sqrt{x^{4}-2\mu_{\lambda}x^{2}+2J_{\lambda}x+C_{\lambda}}}%
=\pm\sqrt{\frac{1}{2}\lambda\rho_{0}^{2}}\left(
\phi-\phi_{0}\right) \equiv\tilde{\phi},\label{197}%
\end{equation}
where $\mu_{\lambda}=\frac{\mu}{\lambda\rho_{0}^{2}},$ \
$J_{\lambda
}=\frac{\left\vert J\right\vert \left(  -1\right)  ^{l}}{\frac{1}{2}%
\lambda\rho_{0}^{3}},$ \ $C_{\lambda}=\frac{C}{\frac{1}{2}\lambda\rho_{0}^{4}%
}$. Let $x_{1},x_{2},x_{3}$, and $x_{4}$ be the roots of equation $x^{4}%
-2\mu_{\lambda}x^{2}+2J_{\lambda}x+C_{\lambda}=0$. From
(\ref{197}) it follows
\[
F\left(
\frac{\sqrt{x_{1}-x_{3}}\sqrt{x_{2}-x}}{\sqrt{x_{1}-x_{2}}\sqrt
{x_{3}-x}},k\right)  =-a\tilde{\phi},
\]
where $a=\frac{1}{2}\sqrt{x_{4}-x_{2}}\sqrt{x_{1}-x_{3}},$ $\
k=\frac{\sqrt
{x_{1}-x_{2}}\sqrt{x_{4}-x_{3}}}{\sqrt{x_{4}-x_{2}}\sqrt{x_{1}-x_{3}}}$.
Here $F\left(  x,k\right) $ is the Jacobi elliptic integral of the
first kind with modulus $k$. Solving the latter equation in $x$,
we arrive at
\begin{equation}
x=\frac{x_{2}\left(  x_{3}-x_{1}\right)  +x_{3}\left(
x_{1}-x_{2}\right)
sn^{2}\left(  a\tilde{\phi},k\right)  }{x_{3}-x_{1}+\left(  x_{1}%
-x_{2}\right)  sn^{2}\left(  a\tilde{\phi},k\right)  },\label{591}%
\end{equation}
where $sn(x,k)$ is the Jacobi elliptic sine with modulus $k$.
(Note that the quantities $x_{1},x_{2},x_{3},x_{4}$ should be
chosen so that $x\geqslant 0$, $\ a^{2}>0$ and $0\leqslant
k^{2}\neq 1$. Otherwise, this unhomogeneous solution does not
exist.) For $k\longrightarrow1$, the period of $sn^{2}(x,k)$ tends
to infinity and the density is independent of the polar angle. For
$k\longrightarrow0$, $sn^{2}%
(x,k)\longrightarrow\sin^{2}(x)$. The position of the extrema of
function $x\left( \phi\right) $ is determined by the quantity
$\xi=\left( x_{1}-x_{2}\right)  \left(  x_{2}-x_{3}\right) \left(
x_{1}-x_{3}\right) $. If $\xi=0$, $x_{\phi}^{\prime}\equiv0$. For
$\xi>0$, the points of minima
$\tilde{\phi}_{l}^{\min}=\frac{K\left(  k\right) }{a}2l,$ \
$x\left(  \tilde{\phi}_{l}^{\min}\right)  =x_{2}$ and maxima
$\tilde{\phi}_{l}^{\max}=\frac{K\left(  k\right) }{a}\left(
2l+1\right) $, $x\left( \tilde{\phi}_{l}^{\max}\right) =x_{1}$,
$l$ being an integer. For $\xi<0$, in these formulae one should
replace $\tilde{\phi}_{l}^{\min
}\rightleftarrows\tilde{\phi}_{l}^{\max}$.

It is seen that, in the general case, the condensate density
$n(\phi)=\rho_{0}^{2}x^{2}\left( \phi\right) $ is a periodic
function (see Fig. 1). The solution obtained allows us to find the
period of density oscillations or, this is the same, the angular
size of a fragment
\begin{equation}
\Delta\phi_{fr}=\frac{2K\left(  k\right)
}{a\sqrt{\frac{1}{2}\lambda
\rho_{0}^{2}}},\label{125}%
\end{equation}
where $K\left(  k\right)$ is the complete elliptic integral. The
number of fragments at the circumference of the ring
$N_{fr}=\frac{2\pi}%
{\Delta\phi_{fr}}$ should be a positive integer. One can, in
principle, find the constant $C=C\left( R\right)$ from this
condition.

Note that the spatially unhomogeneous solutions are also possible
in the lack of the source ($\left\vert J\right\vert =0$). Then
\begin{equation}
n(\phi)=\left(  \rho_{0}x_{1}\right)  ^{2}sn^{2}\left(  \sqrt{\frac{1}%
{2}\lambda\rho_{0}^{2}}\left(  \phi-\phi_{0}\right)
x_{2},k_{0}\right)
,\label{135}%
\end{equation}
where $k_{0}=\frac{x_{1}}{x_{2}},$ \
$x_{1,2}^{2}=\mu_{\lambda}\pm\sqrt
{\mu_{\lambda}^{2}-C_{\lambda}}$.

\begin{figure}[htbp]
\includegraphics[scale=0.5]{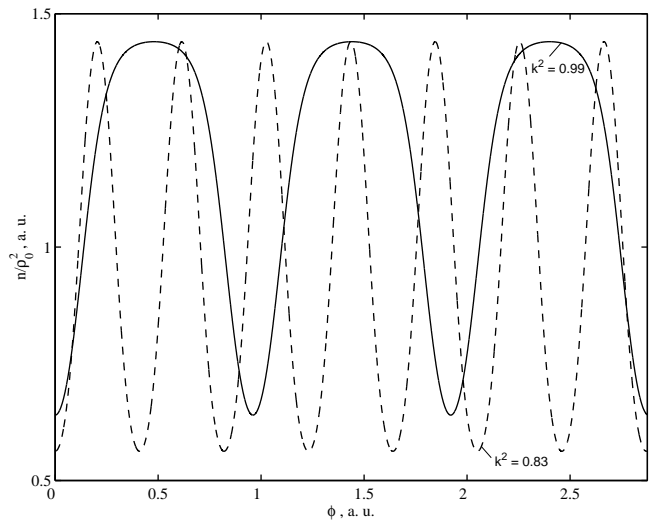}
\caption{The examples of the behavior of the condensate density
$n$ versus polar angle $\phi$ for $x_{1}=1.2,$ $x_{2}=0.8,$
$x_{3}=0.4,$ $x_{4}=1.21$ (solid line) and for $x_{1}=1.2,$
$x_{2}=0.75,$ $x_{3}=0.6,$ $x_{4}=2.1$ (dashed line).}
\end{figure}

\bigskip

\textbf{3. Conclusions and discussion.} To conclude, we have
obtained a set of solutions. All of them, both homogeneous and
unhomogeneous, correspond to the minimum of the energy's
functional of the system, i.e., to the stationary solutions (see
Appendix). This simple theory cannot uniquely predict what a
solution from the set will be realized in fact. However, it is
clear that the spatially homogeneous solution (if it exists at
$J\neq0$) refers to the absolute minimum of energy and
unhomogeneous ones correspond to the steady excited states of the
condensate.

In experiments \cite{1, 2} the IE formation-recombination
processes lead to the condensate density fluctuations in the
radial direction. As is shown in Ref. \cite{12}, at zero
temperature in the cigar-shaped condensate due to the parametric
resonance there takes place one-way transfer of energy from the
transverse (radial) modes to the longitudinal ones. Closing this
"cigar" in a ring, we come to the point that the spatially
non-uniform distribution of the condensate density is nothing else
than a standing wave corresponding to the macroscopically filling
of the longitudinal modes. The consideration of the standing wave
as a steady excited state of the condensate agrees with the
experimental fact that the existing fragmentation starts to vanish
with decreasing the pumping power but for all that the ring radius
would vary negligibly.

Note that the fragmentation of the exciton ring can be explained
in another way, assuming that the IEs are in the condensate state.
Since the "ground" state of the exciton system is degenerate over
the IE spin configurations, it represents a mixture of the
condensates with different coupling constants $\lambda_{ij}>0$ for
the mixture components (here $i,j$ are the indexes of the mixture
components). As is known \cite{13, 14}, if for $i\neq j$ the
condition $\lambda_{ij}^{2}>\lambda_{ii}\lambda_{jj}$ is
fulfilled, the mixture components separate from each other in
space. (On the observation of this effect in atomic condensates
see, e.g., Ref. \cite{15}.) It is possible that such a separation
for the ring geometry is observed in the experiments.

Recently, the observations similar to MOCS ones have been reported
in Ref. \cite{16}. The microscopic derivation (starting with the
electron-hole Hamiltonian) of Eq. (\ref{00}) with $J=0$ is the
same as in Ref. \cite{17}.

\acknowledgments

The authors are thankful to V.S. Babichenko for the suggestion of
the form of the "source-drain" Hamiltonian and helpful comments in
the early stage of the work. We also thank L.A. Maksimov for
valuable discussions. The work is supported by the Russian
Foundation for Basic Research.

\bigskip

\textbf{Appendix.} The total average energy of the system has form
\begin{equation}
E[\rho]=\frac{1}{2\pi}{\int\limits_{0}^{2\pi}}d\phi\left( \left(
\rho_{\phi}^{\prime}\right)
^{2}-\mu\rho^{2}+\frac{1}{2}\lambda\rho ^{4}+2\left\vert
J\right\vert \left(  -1\right)  ^{l}\rho\right)  .
\end{equation}
The minimum of the functional $E[\rho]$ is determined by
Eq.(\ref{2}) and equals
\begin{equation}
E_{\min}[\rho]=\frac{1}{2\pi}{\int\limits_{0}^{2\pi}}d\phi\left(
2\left( \rho_{\phi}^{\prime}\right)  ^{2}-C\right)  ,
\end{equation}
where $C$ is the constant from Eq.(\ref{747}). In the case of the
homogeneous density, involving Eq.(\ref{10}), we arrive at
$E_{\min
}[\rho_{0}]=\mu\rho_{0}^{2}-\frac{3}{2}\lambda\rho_{0}^{4}\equiv-C_{0}$.
This corresponds to the absolute energy minimum provided that
\begin{equation}
\frac{1}{2\pi}{\int\limits_{0}^{2\pi}}d\phi\left(
\rho_{\phi}^{\prime }\right)  ^{2}>\frac{1}{2}\left(
C-C_{0}\right).
\end{equation}

\end{document}